\documentclass{optica-article}

\journal{opticajournal} 

\articletype{Research Article}

\usepackage{color}

\usepackage{amsmath}
\usepackage{bm}

\newcommand{\tcr}[1]{\textcolor{black}{#1}} 

\begin{document}

\title{Experimental investigation of heralded Gaussification of phase-randomized coherent states of light}

\author{Martin Dost\'{a}l, Miroslav Je\v{z}ek, Jarom{\' i}r Fiur{\' a}{\v s}ek\authormark{*}, Jan B\'{\i}lek}

\address{Department of Optics, Faculty of Science, Palack\'{y} University, 17. listopadu 1192/12, 779 00 Olomouc, Czech Republic}

\email{\authormark{*}fiurasek@optics.upol.cz}

\begin{abstract*}
Probabilistic heralded Gaussification of quantum states of light is an important ingredient of protocols for distillation of continuous variable entanglement and squeezing. An elementary step of heralded Gaussification protocol consists of interference  of two copies of the state at a balanced beam splitter, followed by conditioning on outcome of a suitable Gaussian quantum measurement on one output mode. When iterated, the protocol either converges to a Gaussian state, or diverges. Here we report on experimental investigation of the convergence properties of iterative heralded Gaussification.
We experimentally implement two iterations of the protocol, which requires simultaneous processing of four copies of the input state. We utilize the phase-randomized coherent states as the input states, which greatly facilitates the experiment, because these states can be generated deterministically and their mean photon number can easily be tuned. We comprehensively characterize the input and partially Gaussified states by balanced homodyne detection and quantum state tomography.  Our experimental results are in  good agreement with theoretical predictions and they provide new insights into the convergence properties of heralded quantum Gaussification.
\end{abstract*}

\section{Introduction}
Merging quantum states of two optical modes by their interference at a balanced beam splitter followed by projection of one of the output modes onto some specific pure Gaussian state such as the vacuum state represents an important elementary procedure for optical quantum-state manipulation and engineering. Specifically, an iterated version of this protocol applied to many copies of input state $\hat{\rho}$ can Gaussify the state, i.e. convert it into a state with Gaussian Wigner function \cite{Browne2003,Eisert2004,Campbell2012}. Such heralded probabilistic Gaussification was theoretically proposed and investigated \tcr{as a crucial component of} continuous-variable entanglement distillation \cite{Browne2003,Eisert2004}. It is well known that entanglement of Gaussian states cannot be distilled by local Gaussian quantum operations and classical communication \cite{Giedke2002,Eisert2002,Fiurasek2002}. To bypass this no-go theorem, some non-Gaussian operations are required, such as conditional photon subtraction \cite{Opatrny2000,Olivares2003,Bartley2013,Wang2024,Ourjoumtsev2007,Takahashi2010,Kurochkin2014,Dirmeier2020} or single-photon catalysis \cite{Ulanov2015}. \tcr{Since non-Gaussian operations are experimentally much more challenging than Gaussian operations, it is desirable to minimize the number of non-Gaussian operations and apply the non-Gaussian operation to each copy of the input state only once.} Iterative Gaussification protocol can then distill Gaussian entangled state from the de-Gaussified state. \tcr{A more sophisticated version of the protocol, where heralded Gaussification is repeatedly combined with suitable de-Gaussifying operation, can both distill and purify the entanglement such that mixed states are asymptotically distilled to pure two-mode squeezed vacuum state \cite{Fiurasek2010}.}

\begin{figure}[t!]
\centerline{\includegraphics[width=0.8\linewidth]{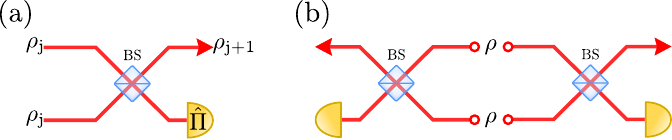}}
\caption{Schematic of a single step of iterative heralded Gaussification protocol. (a) A single-mode version of the protocol. Two copies of the state interfere at a balanced beam splitter BS and success is heralded by observation of a specific outcome of measurement on the second output mode. (b) Two-mode version of the protocol which is suitable for Gaussification of entangled states by local operations and classical communication. \tcr{Two-mode heralded Gaussification can distill continous-variable entanglement via  local Gaussian operations and classical communication provided that the input state $\hat{\rho}_0$ is non-Gaussian.}}
\label{figGscheme}
\end{figure}

Iterative Gaussification can also be utilized to distill entanglement of states that were subject to some de-Gaussifying noise such as phase fluctuations \cite{Hage2008}. A single-mode version of the heralded Gaussification protocol can be used to distill and purify single-mode squeezing \cite{Franzen2006,Marek2007}. \tcr{The heralded Gaussification} protocol does not always converge and its convergence properties were theoretically analyzed in Refs. \cite{Eisert2004,Campbell2012}. The convergence of the protocol and the asymptotic Gaussified state can be partly controlled by the choice of the conditioning measurement on the auxiliary output modes in Fig.~\ref{figGscheme}. A common choice is to perform projection onto vacuum, but projections onto other Gaussian states can also be considered. In particular, with homodyne detection one can condition onto projections onto specific eigenstates of quadrature operator.

Going beyond entanglement and squeezing distillation, the state merging scheme can be used to generate large  Schr\"{o}dinger cat-like states formed by superpositions of coherent states. The scheme can convert two cat-like states with coherent amplitude $\alpha$ into a larger cat-like state with amplitude $\sqrt{2}\alpha$ \cite{Lund2004,Etesse2014,Etesse2015,Sychev2017}.  The `cat breeding' scheme can be iterated to prepare ever larger cat-like states. Moreover, the state merging at a beam splitter can be also used to generate the Gottesman-Kitaev-Preskill states essential for optical quantum computing \cite{Vasconcelos2010,Weigand2018,Konno2024}.
These iterative state-generation schemes become especially powerful if quantum memories are available. The quantum memory significantly enhances the  probability of successful state generation \cite{Cotte2022,Simon2024},  because the partly grown state can be stored in the memory and all elementary input state preparations or merging steps do not need to succeed simultaneously.

In the previous experiments, heralded  Gaussification was mostly limited to a single step of the protocol, with the exception of Ref. \cite{Hage2010} where Gaussification of three copies of a two-mode entangled state was implemented. Multiple iterations of the Gaussification protocol can be emulated by suitable experimental data processing \cite{Abdelkhalek2016}, but this has limited applicability since only the results of specific measurements on the Gaussified state are obtained in this way, and not a physical state.

Here we experimentally implement two iterations of the single-mode heralded Gaussification protocol and systematically investigate its convergence properties. Specifically, we study heralded Gaussification of phase-randomized coherent states. Such states can be \tcr{easily} prepared deterministically, which greatly facilitates the experiment and allows us to experimentally probe two iterations of the protocol. Although non-classical states of light are not generated in our experiment, it nevertheless provides an important insight into the convergence properties of the scheme, which is of general relevance and interest. We utilize single-photon avalanche diodes for heralding, and we condition on non-clicks of these detectors. By contrast, most of the previous experimental implementations of heralded Gaussification employed homodyne detectors for conditioning, which resulted in extra overhead in terms of reduced success probability of the protocol.  

The rest of the paper is organized as follows. In Section 2 we briefly review the theory of heralded Gaussification and then specifically analyze heralded Gaussification of phase-randomized coherent states with emphasis on the convergence properties of the protocol. The experimental setup is described in Section 3 and the experimental results are presented and discussed in Section 4. \tcr{Success probability of heralded Gaussification is discussed in Section 5.} Finally, Section \tcr{6} contains a brief summary and outlook.

\section{Theory}
A single step  of the iterative Gaussification of a single-mode state $\hat{\rho}$ is illustrated in Fig. 1(a). Each step of this protocol is described by the following nonlinear map
\begin{equation}
\hat{\rho}_{j+1}=\frac{\mathrm{Tr}_{2}\left[ \hat{U}_{\mathrm{BS}} \hat{\rho}_j\otimes\hat{\rho}_j\hat{U}_{\mathrm{BS}}^\dagger \hat{I}\otimes \hat{\Pi}\right]}{\mathrm{Tr} \left[ \hat{U}_{\mathrm{BS}} \hat{\rho}_{j}\otimes\hat{\rho}_{j}\hat{U}_{\mathrm{BS}}^\dagger \hat{I}\otimes \hat{\Pi}\right]}.
\label{nonlinearmap}
\end{equation}
Here $\hat{\rho}_{j}$ denotes the partially Gaussified state after $j$th iteration of the protocol, $\hat{\rho}_{0}\equiv\hat{\rho}$  represents the input state, $\hat{U}_{\mathrm{BS}}$ is a unitary operation associated with the balanced beam splitter BS, $\hat{\Pi}$ is a POVM element which specifies the conditioning measurement on mode 2 in Fig. 1(a), and $\mathrm{Tr}_2$ denotes  partial trace over the second mode. \tcr{Note that $p_{\mathrm{succ},j}=\mathrm{Tr} \left[ \hat{U}_{\mathrm{BS}} \hat{\rho}_{j-1}\otimes\hat{\rho}_{j-1}\hat{U}_{\mathrm{BS}}^\dagger \hat{I}\otimes \hat{\Pi}\right]$ represents the probability of success of $j$th elementary step  of the protocol.}

In this work we investigate  Guassification of single-mode states $\hat{\rho}$ which are diagonal in Fock basis. Such states are fully characterized by their photon number distribution $p_n$,
\begin{equation}
\hat{\rho}=\sum_{n=0}^\infty p_n |n\rangle\langle n|.
\label{rhodiagonal}
\end{equation}
Convergence of the Gaussification protocol (\ref{nonlinearmap}) has been theoretically analyzed in Refs. \cite{Eisert2004,Campbell2012} where it was shown that if the protocol converges, then it converges to a Gaussian state.  Note that any Gaussian state $\hat{\rho}_G$ with zero coherent displacement is a fixed point of the nonlinear map (\ref{nonlinearmap}), because 
\begin{equation}
\hat{U}_{\mathrm{BS}} \hat{\rho}_G \otimes \hat{\rho}_G  \hat{U}_{\mathrm{BS}}^\dagger=\hat{\rho}_G \otimes \hat{\rho}_G 
\label{BSGtransform}
\end{equation}
holds for these states.

Let us first consider Gaussification with perfect detectors that can distinguish the presence and absence of photons. Each iteration of the Gaussification protocol succeeds if the second output mode in Fig.~\ref{figGscheme}(a) is projected onto vacuum state, $\hat{\Pi}=|0\rangle\langle 0|$. It is easy to show that for this choice of $\hat{\Pi}$ and the input states of the form (\ref{rhodiagonal}) the map (\ref{nonlinearmap})  preserves the ratio of photon number probabilities $p_1$ and $p_0$. After one step of the protocol the new non normalized photon number probabilities $\tilde{p}_0$  and $\tilde{p}_1$ read
\begin{equation}
\tilde{p}_0=p_0^2,\qquad \tilde{p}_1=\frac{1}{2}p_0p_1+\frac{1}{2}p_1 p_0=p_0 p_1.
\end{equation}
Therefore, we have
\begin{equation}
\frac{\tilde{p}_1}{\tilde{p}_0}=\frac{p_1}{p_0}, 
\end{equation}
as claimed.
Since projection onto the vacuum state is phase-insensitive operation, the Gaussification preserves the Fock-diagonal form (\ref{rhodiagonal}) of the quantum state. The only single-mode Gaussian state with vanishing coherences in Fock basis is the thermal state
\begin{equation}
\hat{\rho}_{\mathrm{th}}=\frac{1}{1+\bar{n}} \sum_{n=0}^\infty \left( \frac{\bar{n}}{1+\bar{n}}\right)^n |n\rangle \langle n|
\label{rhothermal}
\end{equation}
that is fully characterized by its mean photon number $\bar{n}$. It follows that the Gaussification of the Fock-diagonal quantum state (\ref{rhodiagonal}) converges to thermal state if $p_1 <p_0$. The mean photon number \tcr{$\bar{n}_\infty$} of the asymptotic thermal state can be obtained from the condition 
\begin{equation}
\tcr{\frac{\bar{n}_\infty}{\bar{n}_\infty+1}=\frac{p_1}{p_0},}
\end{equation}
which yields
\begin{equation}
\bar{n}_\infty=\frac{p_1}{p_0-p_1}.
\end{equation}

\begin{figure}
\centerline{\includegraphics[width=0.8\linewidth]{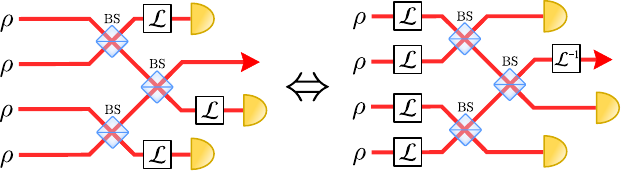}}
\caption{Equivalence between Gaussification with lossy detectors and Gaussification with perfect detection of states transmitted through a lossy channel $\mathcal{L}$. Two iterations of the protocol are illustrated. An inverse operation $\mathcal{L}^{-1}$ needs to be applied to the output state in the right scheme. }
\label{figlosses}
\end{figure}

In the experimental implementation of the Gaussification protocol we utilize single-photon avalanche diodes with limited detection efficiency $\eta$.  Conditioning on non-clicks of such imperfect detectors is equivalent to performing a generalized measurement described by a POVM element
\begin{equation}
\hat{\Pi}_0=\sum_{n=0}^\infty (1-\eta)^n |n\rangle \langle n|.
\end{equation}
Note that this operator is proportional to density matrix of thermal state (\ref{rhothermal}) with mean photon number \tcr{$\bar{n}_\Pi=\eta^{-1}-1$}.
A comprehensive theory of Gaussification with conditioning on general Gaussian measurements was provided in Ref. \cite{Campbell2012} where the convergence criteria  were discussed and an explicit expression for the covariance matrix of the asymptotic Gaussian state was provided. Here we \tcr{first} take a simple approach applicable to scenario, where the heralding measurement is affected by pure losses. As shown in the Supplementary information of Ref. \cite{Grebien2022}, Gaussification with lossy detectors can be equivalently treated as Gaussification of states transmitted over a lossy channel, followed by subsequent inversion of the losses on the asymptotic Gaussian state. This concept is graphically illustrated in Fig.~\ref{figlosses}. 
Let $\mathcal{L}_{\eta}$ denote a lossy channel with transmittance $\eta$.  Consider Gaussification of state $\hat{\rho}_{\eta}=\mathcal{L}_{\eta}(\hat{\rho})$  with perfect detectors and let  $\hat{\sigma}_\eta$ denote the corresponding asymptotic Gaussian state. The actual asymptotic Gaussian state corresponding to Gaussification with imperfect lossy detectors then reads $\mathcal{L}^{-1}_{\eta} (\hat{\sigma}_\eta)$. For thermal state (\ref{rhothermal}), the loss inversion simply rescales the mean photon number as $\bar{n}\rightarrow \bar{n}/\eta$. The vacuum and the single-photon probabilities of the state $ \hat{\rho}_{\eta}$ read
\begin{equation}
p_{0}^\prime=\sum_{n=0}^\infty p_n (1-\eta)^n, \qquad p_{1}^\prime =\sum_{n=1}^\infty n \eta (1-\eta)^{n-1} p_n.
\end{equation}
Mean photon number of the asymptotic thermal state is thus given by
\begin{equation}
\bar{n}_\infty=\frac{1}{\eta}\frac{p_1^\prime}{p_0^\prime-p_1^\prime}
\label{nbarinfinity}
\end{equation}
and the Gaussification converges if $p_0^\prime>p_1^\prime$.
It follows that the convergence of the Gaussification protocol with imperfect detectors depends on the full shape of the photon number distribution $p_n$ of the input state, and not just on the vacuum and single-photon probabilities $p_0$ and $p_1$.
\tcr{The expression (\ref{nbarinfinity}) can be also obtained from the formula for asymptotic covariance matrix $\Gamma_\infty$ of the Gaussified state, that was derived in Ref. \cite{Campbell2012},
\begin{equation}
\tcr{\Gamma_{\infty}=(\Gamma_\Pi-i\Sigma)(\Gamma_{\Pi}-\Gamma_{\sigma})^{-1}(\Gamma_{\Pi}+i\Sigma)-\Gamma_{\Pi}.}
\label{Gammainfty}
\end{equation}
Here  $\Gamma_\Pi$ and $\Gamma_{\sigma}$ denote the covariance matrices of normalized states $\hat{\Pi}_0/\mathrm{Tr}[\Hat{\Pi}_0]$ and $\hat{\sigma}=\hat{\rho}_0\hat{\Pi}_0/\mathrm{Tr}[\hat{\rho}_0\hat{\Pi}_0]$, respectively, and 
\begin{equation}
\Sigma=\left(\begin{array}{cc}0 & 1 \\ -1 & 0 \end{array}\right)
\end{equation}
denotes the symplectic form. For the Fock-diagonal states (\ref{rhodiagonal}), all the covariance matrices appearing in Eq. (\ref{Gammainfty}) are proportional to identity matrix, $\Gamma=(1+2\bar{n})I$. The effective mean photon numbers read $\bar{n}_{\Pi}=(1-\eta)/\eta$ and $\bar{n}_\sigma=(1-\eta)p_1^\prime/(\eta p_0^\prime)$. Since $\Gamma_\infty=(1+2\bar{n}_\infty)I$, the expression (\ref{nbarinfinity}) straightforwardly follows from formula (\ref{Gammainfty}).} 

In the experiment, we compare the heralded  Gaussification that involves conditioning measurement $\hat{\Pi}$ with deterministic unheralded Gaussification. In this latter case, we do not use any conditioning measurement in the protocol, which corresponds to $\hat{\Pi}=\hat{I}$ in Eq. (\ref{nonlinearmap}).  Quadrature operator $\hat{x}$ of the  state after $N$ iterations of deterministic Gaussification  protocol becomes a balanced superposition of $2^N$ quadrature operators of $2^N$ independent copies of the input state, and by the central limit theorem the distribution of $\hat{x}$ will converge to Gaussian distribution in the limit of large $N$. Note that the covariance matrix of the state remains unchanged during deterministic Gaussification so the covariance matrix of the asymptotic Gaussian state is equal to the covariance matrix of the input state. In particular, deterministic Gaussification of an input state diagonal in Fock basis will converge to thermal state with mean photon number $\bar{n}$ equal to the mean photon number of the input state $\hat{\rho}$. \tcr{More generally, deterministic iterative Gaussification converges to a Gaussian state for any input state with zero coherent displacement and finite covariance matrix. This  was  rigorously proven in Ref. \cite{Wolf2006}, where the deterministc Gaussification protocol formed an important part of  derivation of extremality results for Gaussian states.}

In order to quantitatively characterize the Gaussification, we should look at the properties of the quadrature distributions. For states diagonal in Fock basis the probability distributions of all rotated quadratures
\begin{equation}
\hat{x}_\theta=\frac{1}{\sqrt{2}}\left(\hat{a} e^{i\theta}+\hat{a}^\dagger e^{-i\theta}\right)
\end{equation}
are the same and do not depend on $\theta$. Without loss of generality, we can therefore consider distribution $P(x)$ of the amplitude quadrature $\hat{x}=(\hat{a}+\hat{a}^\dagger)/\sqrt{2}$ and analyze the higher moments and cumulants of this distribution. For states diagonal in Fock basis, $P(x)$ is an even function of $x$ and all odd moments of $\hat{x}$ vanish, $\langle \hat{x}^{2k+1}\rangle=0 $. 
We therefore evaluate the excess kurtosis $K$, which can be expressed in terms of second and fourth moments of the quadrature distribution,
\begin{equation}
K=\frac{\langle \hat{x}^4\rangle}{\langle \hat{x}^2 \rangle^2}-3.
\end{equation}
More generally, for distributions with nonvanishing first moment $\langle \hat{x}\rangle=\bar{x}$ we have
\begin{equation}
K=\frac{\langle (\hat{x}-\bar{x})^4\rangle}{\langle (\hat{x}-\bar{x})^2 \rangle^2}-3.
\end{equation}

As an important example relevant for our experiment, consider Gaussification of phase-randomized coherent states, i.e. Fock-diagonal states with Poisson photon number distribution
\begin{equation}
\hat{\rho}= e^{-|\alpha|^2}\sum_{n=0}^\infty \frac{|\alpha|^{2n}}{n!} |n\rangle\langle n|.
\label{rhocohdiagonal}
\end{equation}
For this class of input states the heralded Gaussification protocol converges when
\begin{equation}
|\alpha|^2< \frac{1}{\eta}.
\label{coherentthreshold}
\end{equation}
\tcr{Mean photon number of the asymptotic thermal state can be obtained from Eq. (\ref{nbarinfinity})  and reads}
\begin{equation}
\tcr{\bar{n}_\infty=\frac{|\alpha|^2}{1-\eta|\alpha|^2}.}
\label{nbarcohinfty}
\end{equation}

\begin{figure}[t!]
\centerline{\includegraphics[width=0.9\linewidth]{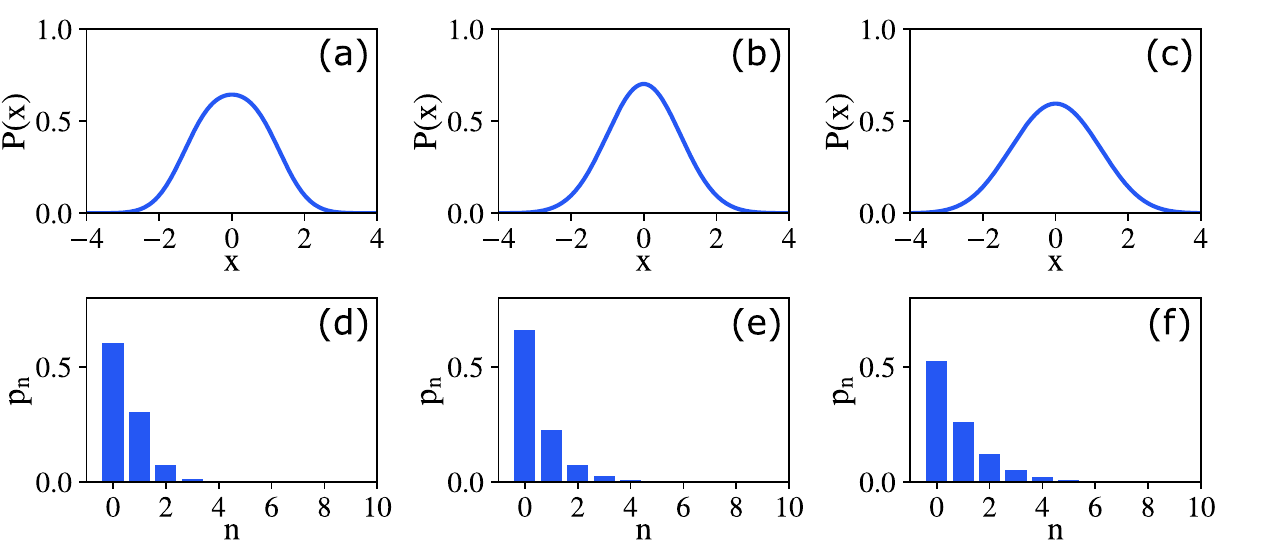}}
\caption{Quadrature distributions $P(x)$ (upper panels) and photon number distributions $p_n$ (lower panels) are plotted for input phase-randomized coherent state (\ref{rhocohdiagonal}) with $|\alpha|^2=0.75$ (a,d), for state obtained after three iterations of deterministic Gaussification (b,e), and for state obtained by three iterations of conditional heralded Gaussification (c,f). Perfect detectors with $\eta=1$ are assumed.}
\label{figtheorybelow}
\end{figure}
\begin{figure}
\centerline{\includegraphics[width=0.9\linewidth]{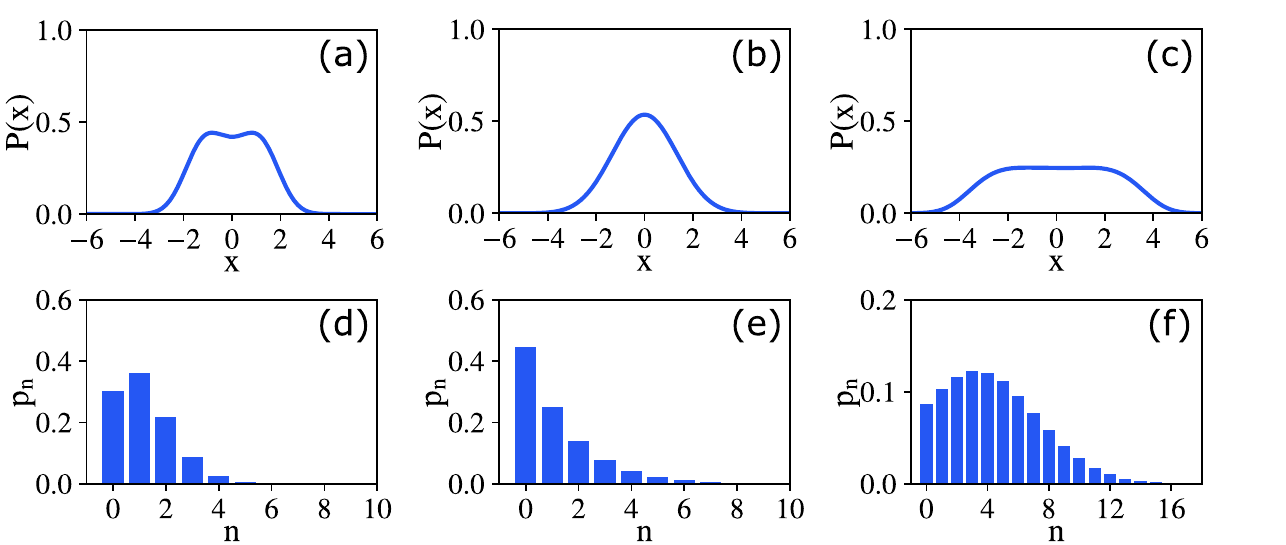}}
\caption{This figure shows the same results as Fig.~\ref{figtheorybelow} but $|\alpha|^2=1.25$, hence the mean photon number of the input state is above the convergence threshold  of the conditional Gaussification protocol.}
\label{figtheoryabove}
\end{figure}

The difference between the behavior of the heralded Gaussification below and above the convergence threshold is illustrated in Figs. \ref{figtheorybelow} and \ref{figtheoryabove}. These figures contain plots of quadrature distributions $P(x)$ and the photon number distributions $p_n$  for input dephased coherent state, state after three iterations of deterministic Gaussification, and state after three iterations of heralded conditional Gaussification. The results are plotted for  $|\alpha|^2=0.75$ and $|\alpha|^2=1.25$, respectively. Perfect photon detectors with unit detection efficiency are assumed. The deterministic Gaussification protocol converges to thermal state in both cases. On the other hand, the heralded  Gaussification converges to thermal state only for the first considered example $|\alpha|^2=0.75$. When the input mean photon number exceeds the threshold value $1$, the conditional Gaussification does not converge, mean photon number of the state increases after each iteration of the protocol, and the peak of the photon number distribution $p_n$ shifts to larger $n$.

\begin{figure}[t!]
\centerline{\includegraphics[width=\linewidth]{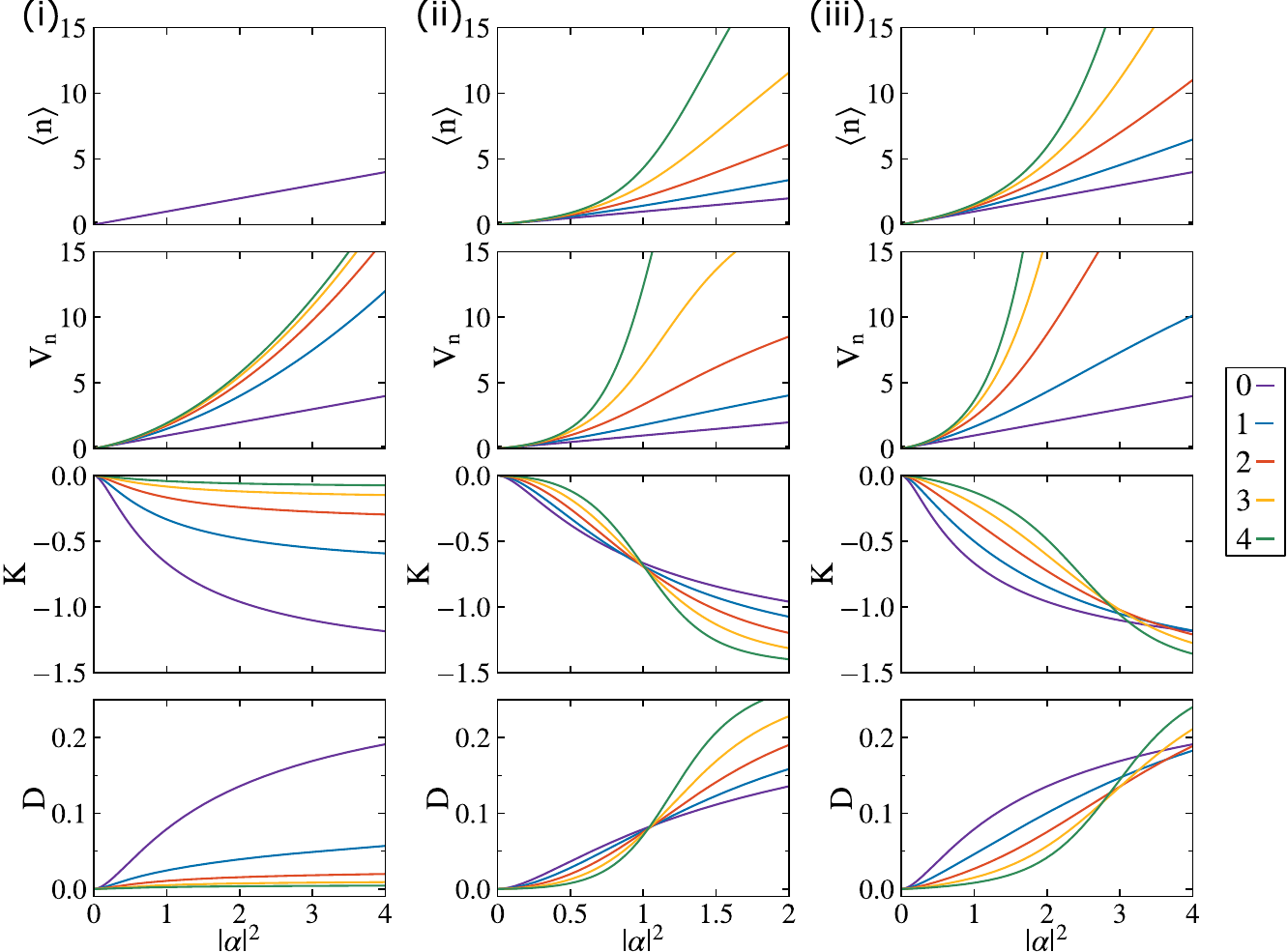}}
\caption{Gaussification of phase-randomized coherent states. The mean photon number $\bar{n}$, photon number variance $V_n=\langle(\Delta \hat{n})^2\rangle$, the excess kurtosis $K$, \tcr{and the statistical distance $D$} are plotted as functions of the input mean photon number $|\alpha|^2$ for three scenarios: (i) deterministic Gaussification (left column), (ii) conditional heralded Gaussificatiion with perfect detectors (middle column), and (iii) conditional heralded Gaussification with imperfect detectors with $\eta_{\mathrm{APD}}=0.4$ (right column). Line colors specify  the number of iterations as indicated in the side panel.}
\label{figkurtosistheory}
\end{figure}

To further quantitatively characterize the performance of the Gaussification protocols, we plot in Fig.~\ref{figkurtosistheory} the mean photon number $\bar{n}$, the photon number variance $V_n=\langle (\Delta \hat{n})^2\rangle$  and the excess kurtosis $K$  for several iterations of the protocol  as functions of the input mean photon number $|\alpha|^2$. To better connect the theoretical analysis to experiment, we consider in this figure also Gaussification with inefficient single-photon detectors with detection efficiency $\eta=0.4$. Both Fig.~4 and 
Fig.~5 illustrate the qualitative difference between deterministic and heralded Gaussification above the convergence threshold of the heralded scheme.

\begin{figure}[t!]
\centerline{\includegraphics[width=\linewidth]{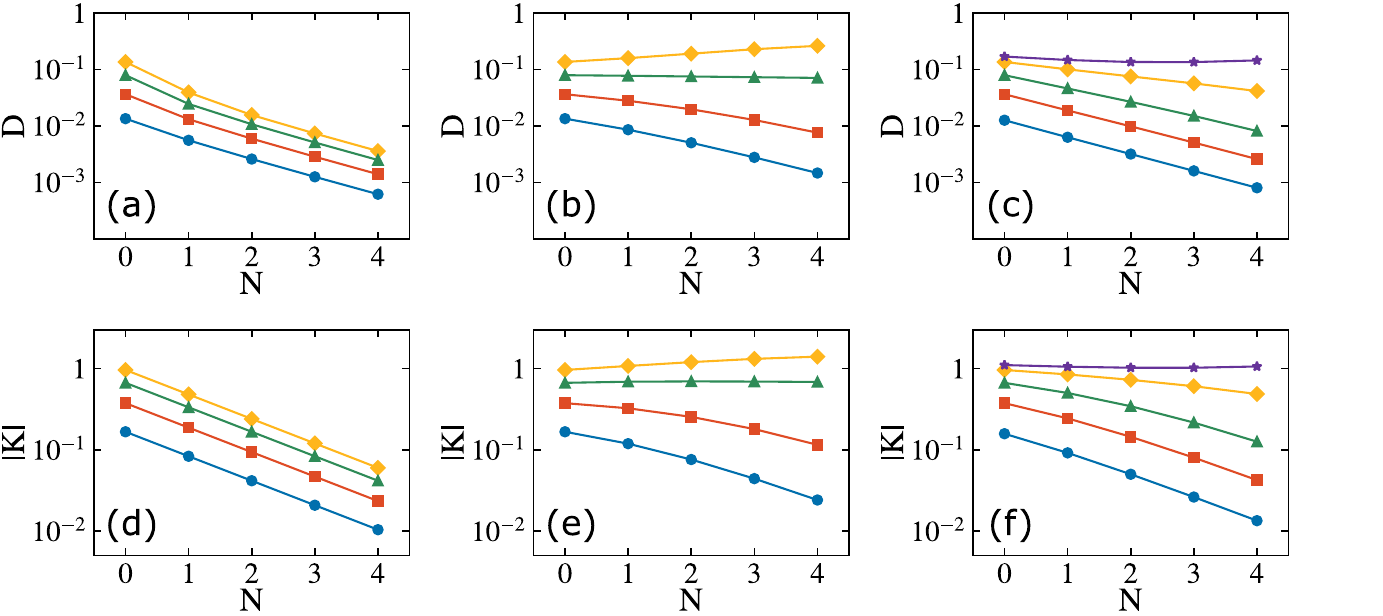}}
\caption{\tcr{Convergence of Gaussification protocols. The absolute value of excess kurtosis $|K|$ and the statistical distance $D$ are plotted in dependence on the number of iterations $N$ of the Gaussification protocol. Results are shown for deterministic Gausssification (a,d), heralded Gaussification with perfect detectors (b,e) and heralded Gaussification with imperfect detectors with detection efficiency $\eta=0.4$ (c,f). Color represents the value of $|\alpha|^2$ as follows: $|\alpha|^2=0.25$ (blue circles), $|\alpha|^2=0.5$ (red squares), $|\alpha|^2=1$ (green triangles),  $|\alpha|^2=2$ (yellow diamonds), $|\alpha|^2=3$ (purple stars). The lines serve to guide the eye.}} 
\label{figconvergence}
\end{figure}

\tcr{Besides looking at higher order cumulants of the quadrature distribution $P(x)$, we can directly consider the distance of a given quantum state from Gaussian state. The non-Gaussian character of a quantum state can be for instance quantified by a suitably normalized Hilbert-Schmidt distance of the state from a Gaussian state with the same coherent displacement and covariance matrix \cite{Genoni2007}. In our experiment, we directly sample the phase-averaged quadrature distribution $P(x)$ which fully characterizes the Fock-diagonal states (\ref{rhodiagonal}) we deal with. It is therefore natural to quantify the difference between a given state and a Gaussian state via the difference between the sampled quadrature distribution $P(x)$ and a Gaussian distribution $P_G(x)$ with the same mean and variance. We utilize the statistical distance $D$ for this purpose, which is defined as follows, 
\begin{equation}
D=\frac{1}{2}\int_{-\infty}^\infty |P(x)-P_G(x)| dx.
\end{equation}
The parameter $D$ is  plotted  in Fig. 5 for both heralded and deterministic Gaussification protocols. Observe that for the states considered in this work the behavior of $D$ is qualitatively fully equivalent to the behavior of the excess kurtosis $K$. For deterministic Gaussification, the protocol converges to Gaussian state irrespective of the value of $|\alpha|$. On the other hand, the heralded Gaussification converges only for sufficiently small initial mean photon number, while both $K$ and $D$ can  increase with the increasing number of iterations above the convergence threshold. }

\tcr{To better visualize the speed of convergence of Gaussification of phase-randomized coherent states, we plot in Fig.~\ref{figconvergence} the excess kurtosis $K$ and the statistical distance $D$ in dependence on the total number of iterations $N$ of the protocol for several chosen values of $|\alpha|^2$. In case of deterministic Gaussification, the convergence is exponentially fast  in $N$ and the convergence rate essentially does not depend on $|\alpha|^2$. For the excess kurtosis $K$ it can be rigorously proven that after $N$ iterations the excess kurtosis will read $K_N=K_0/2^N$, where $2^N$ is the total number of state copies wich are averaged in the protocol and $K_0$ is the excess kurtosis of the initial state. By contrast, convergence rate of heralded Gaussification slows down with increasing $|\alpha|$ and vanishes as we approach the threshold. }

\section{Experiment}

\begin{figure}[b]
\centerline{\includegraphics[width=0.9\linewidth]{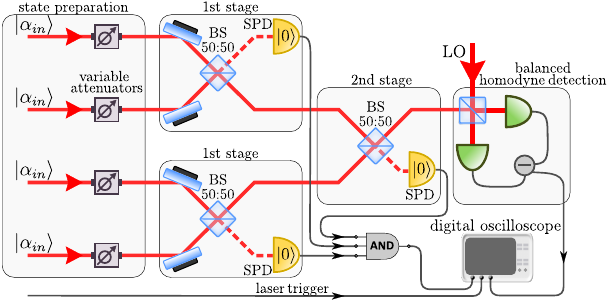}}
\caption{Scheme of the experimental setup for Gaussification of phase-randomized coherent states with fundamental blocks denoted, namely the input state preparation, pair-wise interference of optical beams, and homodyne detection. }
\label{figexperimentalsetup}
\end{figure}

The experimental setup is depicted in Fig.~\ref{figexperimentalsetup}. The optical signals are derived from a gain-switched laser diode producing 5 ns  pulses with a repetition frequency of 0.9 MHz at a wavelength of 808 nm. The laser is split into four signal branches $|\alpha_{\mathrm{in}}\rangle$  and strong local oscillator (LO) for homodyne detection. The input signal states are prepared with the desired intensity using variable attenuators. \tcr{We generate phase-randomized coherent states by modulating the phases of all input states with coprime frequencies and acquiring data over many modulation periods}. Input states are pair-wise interfered in Michelson interferometers, denoted as the 1st stage, while the phase of each input state is independently controlled and randomized via mirrors attached to the piezo actuators. The interferometers' off-axis configuration gives access to both output ports. One output from each interferometer is routed to the 2nd interferometric stage realized by a 50:50 beam-splitter (BS). From the 2nd stage, one output is interfered with the local oscillator and measured by a homodyne detector. The remaining outputs of the 1st and 2nd stages marked with dashed red lines in the scheme are coupled to single-mode optical fibers and routed to single-photon avalanche diodes which serve as single photon detectors (SPD) that can distinguish the presence and absence of photons. Our home-built homodyne detector, based on a charge-sensitive amplifier, reaches a signal-to-noise ratio of more than 22 dB (with 10 $\mu$W LO power) and a bandwidth of 1 MHz.

\tcr{Both the deterministic Gaussification and the heralded probabilistic Gaussification are implemented with the same experimental setup and the difference is in the heralding signal. When we perform deterministic Gaussification, data acquisition is triggered solely by the electronic laser trigger that signals the presence of laser pulse at the setup inputs, and we do not use signals from the single-photon detectors SPD.  In deterministic Gaussification, all events are thus accepted, irrespective of the measurement outcomes of SPDs. By contrast, heralded Gaussification involves conditioning on non-clicks of SPDs. When we implement one iteration of the heralded Gaussification, we condition on non-click of one SPD, and when we implement two iterations of the protocol, we condition on simultaneous non-clicks of all three SPDs in the setup alongside the laser trigger. We block three input modes of the setup to measure the input phase-randomized coherent state. We block two inputs to realize single iteration of Gaussification. Finally, all four input modes in Fig.~\ref{figexperimentalsetup} are used to realize two iterations of the protocol. 
}

We  process the experimental data by two approaches. First, we directly estimate moments of quadrature distributions and photon number distributions by averaging suitable pattern functions over the sampled quadrature statistics \cite{Welsch1999,DAriano2003}. Second, we reconstruct the photon number distributions of the measured quantum states by maximum-likelihood estimation algorithm \cite{Jezek2003,Hradil2004,Lvovsky2004}. 
The efficiency of the  homodyne detector is $\eta_{\mathrm{BHD}}=0.65$. Additionally, due to the construction of our setup the measurements of input states (states after first stage of Gaussification) are influenced by additional losses of $75\%$  ($50\%$) because the signal propagates through one or two balanced beam splitters. In what follows we denote by $\eta_H$ the overall detection efficiency of the homodyne detector which includes also the losses caused by transmissions through the balanced beam splitters.

As noted in the previous Section, for the  fully phase randomized states (\ref{rhodiagonal}) the quadrature distributions of all the rotated quadratures $\hat{x}_{\theta}$ are the same. In what follows we therefore use the symbol $\hat{x}$ to denote the quadrature and drop the angle $\theta$.
A lossy homodyne detector measures the statistics of the quadrature 
\begin{equation}
\hat{x}_{M}=\sqrt{\eta_H}\hat{x}+\sqrt{1-\eta_H}\hat{x}_{\mathrm{vac}},
\end{equation}
where $\hat{x}_{\mathrm{vac}}$ denotes a quadrature operator of an auxiliary vacuum mode. Various quantities of interest can be determined by averaging suitable loss-compensating pattern functions over the experimentally sampled quadrature statistics  $P_M(x_M)$ \cite{Richter1996,Richter2000,Fiurasek2001}.  In particular, moments of the quadrature operator $\hat{x}$ can be determined as follows,
\begin{equation}
\left\langle\hat{x}^k \right\rangle = \left(\frac{\sqrt{1-\eta_H}}{2\sqrt{\eta_H}}\right)^k \int_{-\infty} ^{\infty} H_k\left( \frac{x_M}{\sqrt{1-\eta_H}}\right) P_M(x_M)   dx_M.
\label{samplingxmoment}
\end{equation}
Here $H_k(x)$ denotes Hermite polynomial of degree $k$. Similarly, the normally ordered moments of photon number distribution of the measured state can be expressed as
\begin{equation}
\langle :\hat{n}^k\!:\rangle=\langle \hat{a}^{\dagger k} \hat{a}^k\rangle= \frac{1}{\eta_H^k} \frac{k! k!}{2^k (2k)!}\int_{-\infty}^\infty  H_{2k}(x_M) P_M(x_M) d x_M.
\label{samplingnnormal}
\end{equation}
This formula reflects the well-known fact that the normally ordered moments of photon number distribution scale under losses as 
\begin{equation}
\langle \hat{a}^{\dagger k} \hat{a}^k\rangle \rightarrow \eta_H^k \langle \hat{a}^{\dagger k} \hat{a}^k\rangle.
\end{equation}
In what follows we focus in particular on the mean photon number $\langle \hat{n}\rangle=\langle \hat{a}^\dagger \hat{a}\rangle$ and the photon number variance 
\begin{equation}
\langle (\Delta \hat{n})^2\rangle= \langle \hat{a}^{\dagger 2}\hat{a}^2\rangle+\langle \hat{a}^\dagger \hat{a}\rangle- (\langle \hat{a}^\dagger \hat{a} \rangle)^2.
\end{equation}
In experimental data processing, the integrals in Eqs. (\ref{samplingxmoment}) and (\ref{samplingnnormal}) are replaced by normalized discrete sums over all sampled quadrature values.

\section{Experimental results}

We have measured the quadrature distributions $P(x)$ of the input states, states after one round of Gaussification, and states after two rounds of Gaussification  for phase-randomized coherent states with various input mean photon numbers. Due to intensity fluctuations of the laser diode that serves as a source of coherent light in our experiment, the generated input states $\hat{\rho}$ are not exactly the phase-randomized coherent states in Eq. (\ref{rhocohdiagonal}) with Poisson photon number distribution, but they exhibit slightly super-Poissonian photon number statistics. 

\begin{figure}
\centerline{\includegraphics[width=\linewidth]{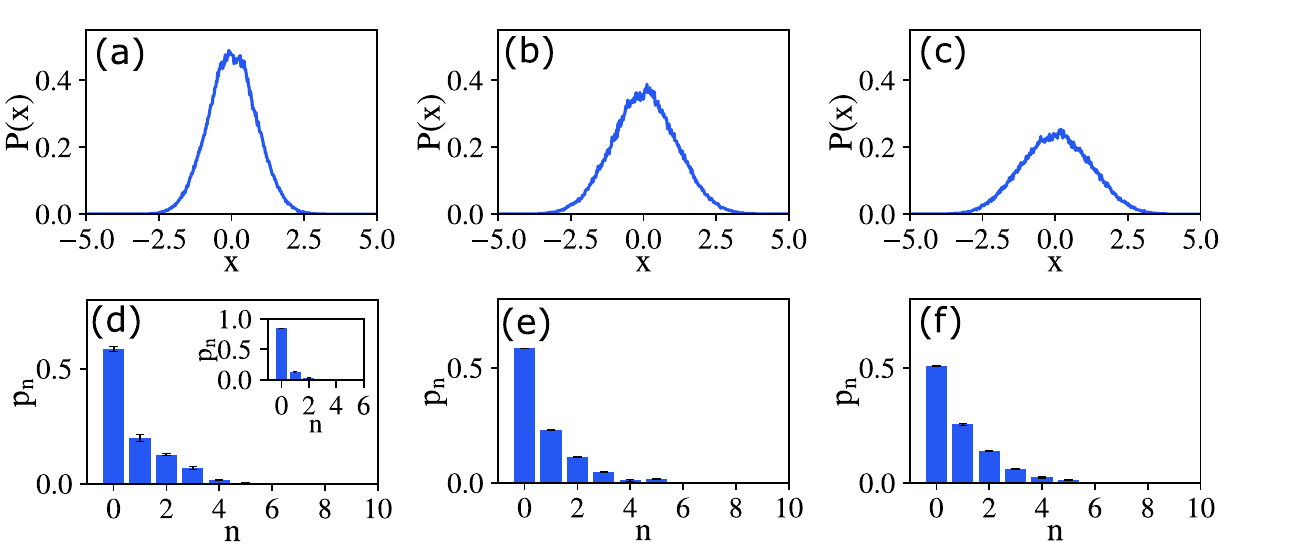}}
\caption{Experimentally measured quadrature distributions $P(x)$ (upper panels) and the corresponding reconstructed photon number distributions $p_n$ (lower panels) are plotted for input state  with $\langle \hat{n}\rangle_{\mathrm{in}}=1.13\pm0.01$ (a,d), for state obtained after two iterations of deterministic Gaussification (b,e), and for state obtained by two iterations of the conditional heralded Gaussification (c,f). Note that the measurement of the quadrature distribution of the input state is affected by $75\%$ losses due to transmission through two balanced beam splitters in the experimental setup. The inset in panel (d) shows the photon number distribution reconstructed without any loss compensation, while the main panel displays  photon number  distribution obtained via loss-compensating quantum state tomography.}
\label{figexpnsmall}
\end{figure}

\begin{figure}
\centerline{\includegraphics[width=\linewidth]{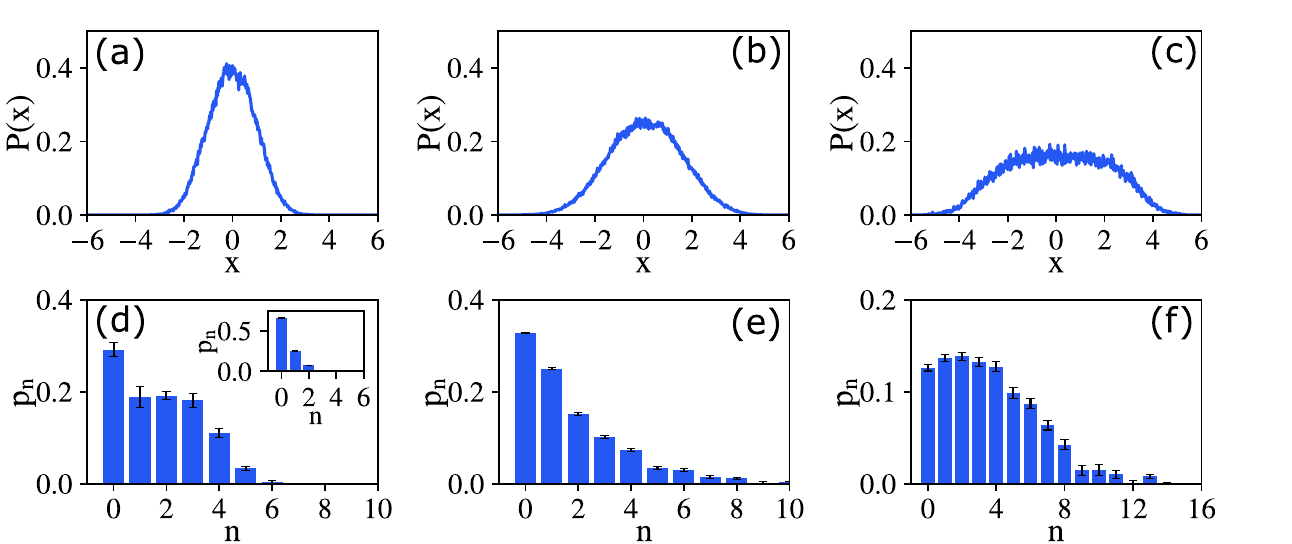}}
\caption{This figure shows the same results as Fig.~\ref{figexpnsmall} but for $\langle \hat{n}\rangle_{\mathrm{in}}=2.69\pm 0.02$.}
\label{figexpnlarge}
\end{figure}

Explicit examples of the behavior of  heralded and deterministic Gaussification protocols for small and large mean photon number of the input state are illustrated in Figs. \ref{figexpnsmall} and \ref{figexpnlarge}. There we plot the experimentally sampled phase-averaged quadrature distributions $P(x)$ as well as the reconstructed photon number distributions $p_n$ for two different input mean photon numbers $\langle \hat{n}\rangle=1.13\pm 0.01$ and $\langle\hat{n}\rangle=2.69\pm 0.02$. Since the single-photon detectors utilized in the experiment exhibit overall detection efficiency of about $\eta=0.40$, the first example is well below the convergence threshold of heralded Gaussification, while the second example is close to the threshold. See Eq. (\ref{coherentthreshold}) which  predicts the threshold $\langle \hat{n}\rangle=2.5$ for the phase-randomized coherent states and $\eta=0.40$.  Well below the threshold, both deterministic and heralded Gaussification converge to thermal state and the fidelities of the reconstructed photon number distributions in Fig. \ref{figexpnsmall}(e,f) with the thermal state distribution with the same mean photon number read $F_{\mathrm{det}}=0.992\pm 0.002$ and $F_{\mathrm{her}}=0.993\pm0.002$, respectively. By contrast, in the second example given in Fig.~\ref{figexpnlarge} the photon number distribution after two rounds of heralded Gaussification is far from that of a thermal  state and the quadrature distribution $P(x)$ becomes very broad and flat. The fidelity of deterministically Gaussified state in Fig.~\ref{figexpnlarge}(e) with a thermal state remains very high, $F_{\mathrm{det}}=0.994\pm 0.002$, while the fidelity of the heralded state in Fig.~\ref{figexpnlarge}(f) with a thermal state is much smaller, $ F_{\mathrm{her}}=0.925\pm 0.006$. 

Statistical uncertainty of the reconstructed photon number distributions was determined by Monte Carlo simulations. Assuming that the photon number distributions plotted in Figs.~\ref{figexpnsmall} and \ref{figexpnlarge}  are the true distributions, we have simulated 100 times the whole homodyne measurement. For each simulated measurement we have reconstructed the photon number distribution from the simulated homodyne data. Finally, we have determined the statistical uncertainties of $p_n$ from the resulting ensemble of reconstructed photon number distributions. The error bars plotted in panels (d)-(f) of Figs.~\ref{figexpnsmall} and \ref{figexpnlarge} represent one standard deviation.

The observed experimental results are in very good qualitative agreement with the theoretical predictions in Figs.~\ref{figtheorybelow} and \ref{figtheoryabove}. However, since our input states are not exactly the phase-randomized coherent states, it is difficult to make exact quantitative comparison with the theoretical model discussed in Section 2. Note also that the reconstruction of the input photon number distribution of the input state is affected by $75\%$ losses which enhances errors and uncertainty of the estimation. By contrast, the reconstruction of the photon number distributions of states after two rounds of Gaussification does not suffer from such losses and is affected only by efficiency of the homodyne detector $\eta_{\mathrm{BHD}}=0.65$. We do not compensate for the homodyne detector efficiency $\eta_{\mathrm{BHD}}$ in Figs. \ref{figexpnsmall} and \ref{figexpnlarge}.

\tcr{Besides the laser power fluctuations and other experimental imperfections such as limited interference visibility, imperfect phase randomization \cite{Cao2020} could possibly affect the experiment. However, the experimental data indicate that the phase randomization works reasonably well. The experimentally sampled quadrature distributions $P(x)$ exhibit the expected symmetric shape and the experimentally observed mean quadrature values are all less than $0.09$ in absolute value. However, it should be noted that the states measured by the homodyne detector are subjected to additional phase randomization because the relative phase between the signal and the local oscillator is not locked. We can therefore also consider the evolution of mean photon number during deterministic Gaussification. Presence of a significant coherent component in input states  would lead to interference effects resulting in changes of the mean photon number of the deterministically Gausisfied state. With the exception of a single point at $|\alpha|^2=2.04$ in Fig. 10(a), the mean photon numbers remain practically constant or change only minimally  during deterministic Gaussification. Moreover, the obseved slight changes of mean photon number for certain $|\alpha|^2$ can be caused by other imperfections and fluctuations, becuase the measurements are performed sequentially.}

\begin{figure}
\centerline{\includegraphics[width=\linewidth]{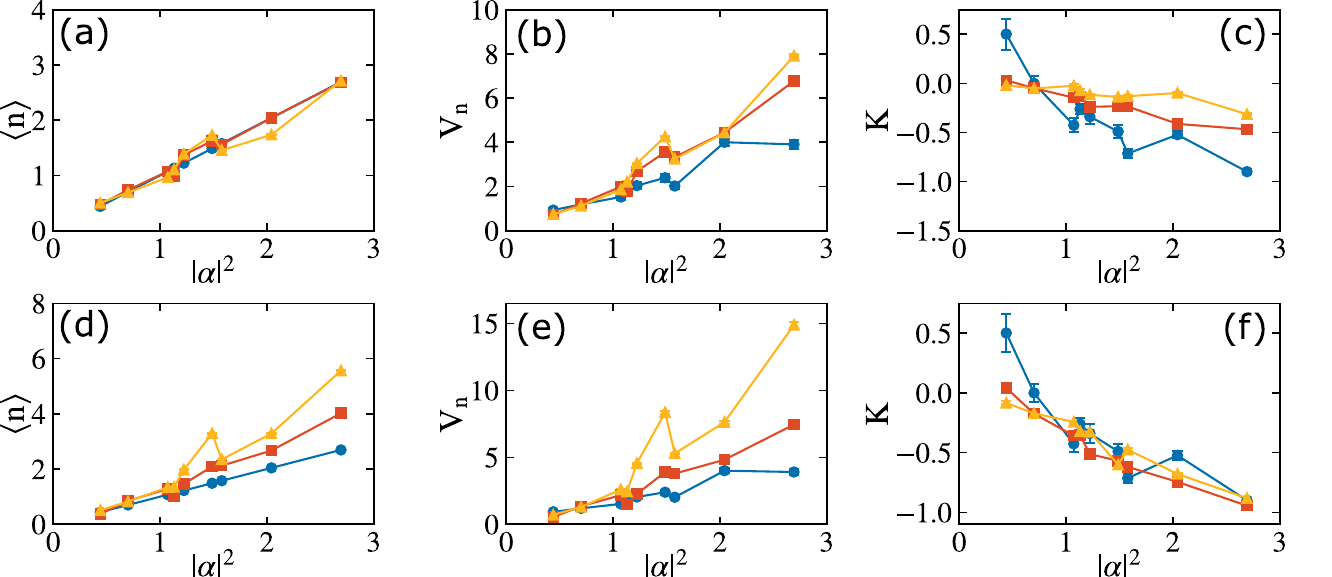}}
\caption{Experimentally determined  mean photon numbers $\langle \hat{n}\rangle$, photon number variances $V_n=\langle (\Delta \hat{n})^2\rangle$ and the excess kurtosis $K$ of quadrature distribution of the input states (blue dots), states after one iteration of Gaussification (red squares), and states after two iterations of Gaussification (yellow triangles). Upper panels display results for deterministic Gaussification, while the lower panels show results for heralded probabilistic Gaussification. Symbols represent experimental results, the lines serve to guide the eye. Statistical uncertainties of the estimated quantities are indicated by error bars that represent one standard deviation. Note that for certain data points the error bars are smaller than the symbol size.} 
\label{figexpK}
\end{figure}

To fully explore the recorded experimental data, we utilize the loss-compensating pattern functions (\ref{samplingxmoment}) and (\ref{samplingnnormal}) to estimate quadrature moments and normally ordered photon number moments from the measured quadrature distributions of the input states and states after one and two rounds of Gaussification. To obtain fully comparable data, we now compensate for the inefficient homodyne detection as well as for the additional losses imposed by the beam splitters. From the reconstructed moments  we calculate the photon number variances and the excess kurtosis $K$. The obtained results are plotted in Fig.~\ref{figexpK} for both deterministic and probabilistic heralded Gaussifications. Symbols represent experimental data and the lines serve to guide the eye. As before, the results are in very good qualitative agreement with the theoretical predictions, c.f Fig.~\ref{figkurtosistheory}. 

For deterministic Gaussification, we observe that the mean photon number remains constant and does not change with the iterations of the protocol. By contrast, in case of heralded Gaussifcation the mean photon number increases at every step of the protocol. The photon number variance grows much faster for heralded Gaussification than for the deterministic one. Most importantly, we clearly observe the expected difference in the behavior of the excess kurtosis $K$. Deterministic unheralded Gaussification quickly converges to Gaussian thermal state irrespective of the input mean photon number and each iteration of the protocol reduces $K$. In case of heralded probabilistic Gaussification we observe significant reduction of the excess kurtosis only for the lowest  input mean photon numbers while for larger $\bar{n}$ the reduction of $K$ becomes only marginal or $K$ even slightly increases with the number of elementary steps of the protocol.

\begin{figure}[t]
\centerline{\includegraphics[width=\linewidth]{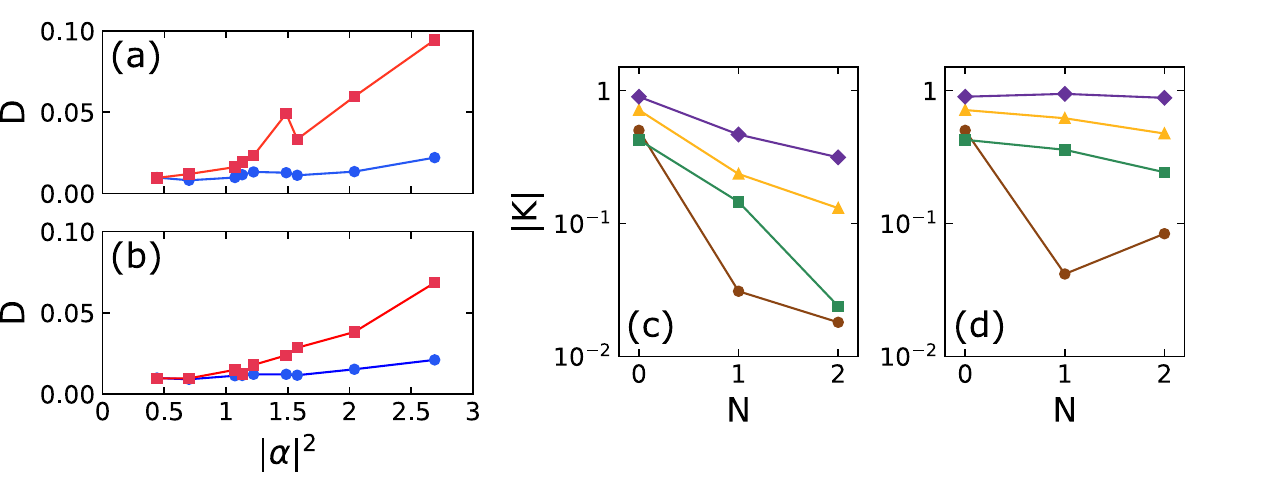}}
\caption{\tcr{Experimentally determined statistical distance $D$ generated by one (a) and two (b) iterations of Gaussification of phase-randomized coherent states is compared for deterministic Gaussification (blue dots) and heralded Gaussification (red squares). The figure shows also the dependence of the experimentally determined excess kurtosis $K$ on the number of iterations of the protocol $N$ for deterministic (c) and heralded (d) Gaussification and for four different input mean photon numbers $0.44$ (brown dots)  $1.07$ (green squares), $1.58$ (blue triangles), and  $2.69$ (purple diamonds). Statistical error bars are smaller than the size of the symbols.}} 
\label{figexpconvergence}
\end{figure}

\tcr{The experimental convergence of  Gaussification  is further illustrated in Fig.~\ref{figexpconvergence} where we plot the dependence of $|K|$ on the number of iterations of the protocol  for several chosen values of input  mean photon number. The advantage of the excess kurtosis $K$ is that it can be reliably and consistently determined for all stages of the experimental Gaussification using the loss-compensating sampling functions (\ref{samplingxmoment}). For completeness we have also evaluated the statistical distance $D$ of the experimentally sampled quadrature distributions $P(x)$ from Gaussian distributions $P_G(x)$ with the same first and second moments. Since the input states are measured with additional losses of $75\%$ and the states after one round of Gaussification are measured with additional $50\%$ losses, the statistical distances $D$ that we evaluate for various stages of the protocol are not directly comparable. For a given fixed number of iterations we can nevertheless consistently compare the results of heralded and deterministic Gaussification, see  Fig.~\ref{figexpconvergence}(a,b). To calculate $D$ from the experimental data, we have discretized the distribution by binning into $200$ bins. We have also binned the corresponding reference Gaussian distribution $P_G(x)$ to ensure correct comparison. The number of bins was chosen such as to faithfully capture the shape of the distributions  while at the same time ensuring that the statistical fluctuations of the number of samples in each bin are not too large. All the experimental results consistently and reliably demonstrate the differences between deterministic thermalization and heralded conditional Gaussification. }

Due to experimental imperfections, the input state with the lowest mean photon number exhibits positive excess kurtosis $K$, although theory predicts negative $K$ for the phase-randomized coherent states. Positive excess kurtosis indicates that we generate quantum state with $g^{(2)}$ factor larger than $2$, where
\begin{equation}
g^{(2)}= \frac{\langle \hat{a}^{\dagger 2} \hat{a}^2\rangle}{\langle \hat{a}^\dagger \hat{a} \rangle^2}.
\end{equation}
In fact, it turns out that for the Fock diagonal states $\hat{\rho}$ the  excess kurtosis K of quadrature distribution is simply related to the $g^{(2)}$ factor,
\begin{equation}
K=\frac{6 \langle \hat{n}\rangle^2}{(2\langle \hat{n}\rangle+1)^2}\left( g^{(2)}-2\right).
\end{equation}
We have  $g^{(2)}=2$ for thermal states and $g^{(2)}=1$  for phase-randomized coherent states. In the limit of low mean photon number the experimental imperfections and source fluctuations become particularly significant and lead to input state with $g^{(2)}>2$. For the second lowest experimental  mean photon number we generate input state with excess kurtosis very close to $0$,  which is preserved throughout the Gaussification.

\section{\tcr{Success probability of heralded Gaussification}}

\tcr{In this section we discuss the success probability of heralded Gaussification and compare theoretical predictions with experimental results. Let $p_{\mathrm{tot},N}$ denote the total success probability of  $N$ iterations of the protocol, i.e. the probability that the heralded Gaussification scheme converts $2^N$ input copies of the state into a single copy of partially Gaussified state.  Since all elementary steps of Gaussification must succeed simultaneously, we can express $p_{\mathrm{tot}, N}$ in terms of probabilities $p_{\mathrm{succ},j}$ of success of each elementary step of iterative Gaussification as fololows,
\begin{equation}
p_{\mathrm{tot}, N}=\prod_{j=1}^N p_{\mathrm{succ},j}^{2^{N-j}}.
\end{equation}
The success probability $p_{\mathrm{tot}, N}$ decreases exponentially with the number of processed copies, because the success is heralded by simultaneous non-clicks of all $2^N-1$ detectors in the setup. This scaling could be improved by using quantum memories, similarly to quantum repeater architectures, but this would make the experiment much more challenging and sensitive to any imperfections of the utilized quantum memories. }

\begin{figure}[t]
\centerline{\includegraphics[width=0.85\linewidth]{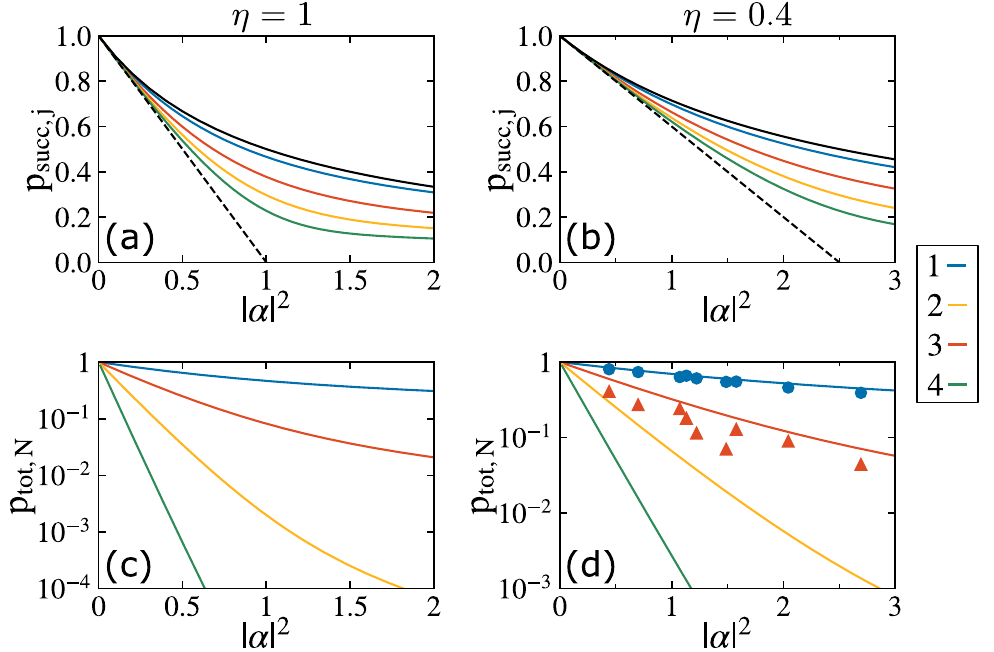}}
\caption{\tcr{Success probability of heralded Gaussification. The success probabilities $p_{\mathrm{succ},j}$ of single iteration of the protocol  (a,b) as well as the total probabilities  $p_{\mathrm{tot},N}$  of $N$ iterations of the protocol (c,d) are plotted as functions of $|\alpha|^2$  for $\eta=1$ (a,c) and $\eta=0.4$ (b,d). The line colors indicate the number of iterations. Black solid and dashed lines in panels (a,b) represent the upper and lower bounds on $p_{\mathrm{succ},j}$,  specified in the main text. Experimentally determined probabilities $p_{\mathrm{tot},N}$ are plotted in panel (d) as blue dots ($N=1$) and red triangles ($N=2$). Statistical error bars are smaller than the size of the symbols.} }
\label{figprobability}
\end{figure}

\tcr{In Fig.~\ref{figprobability} we plot the theoretically calculated probabilities $p_{\mathrm{succ},j}$ and $p_{\mathrm{tot},N}$ for heralded Gaussification of phase-randomized coherent states (\ref{rhocohdiagonal}). We consider both ideal detectors with unit detection efficiency as well as realistic detectors with total detection efficiency $\eta=0.4$ which ocrresponds to our experiment.  We also plot in Fig.~\ref{figprobability}(d) the experimental probabilities $p_{\mathrm{tot},1}$ and $p_{\mathrm{tot},2}$ which were estimated as the ratio of the number of successful heralding events and the total number of laser pulses sent to the setup during the measurement time. We observe very good agreement between experiment and theory for $p_{\mathrm{tot},1}$. The discrepancies between the theory and experiment are slighly larger for $p_{\mathrm{tot},2}$ due to experimental imperfections, because the effective experimental configuration for two iterations of the protocol is much more complicated than for one iteration. Additionally, the theoretical predictions become more sensitive to precise estimation of $\eta$. Nevertheless, the qualitative agreement with theory is still reasonably good and $p_{\mathrm{tot},2}$ follows the theoretically predicted trend.}

\tcr{If the heralded Gaussification of phase-randomized coherent states converges, then we can expect that the photon number statistics of the partially Gaussified states are somewhere in-between the Poisson distribution and Bose-Einstein distribution. Let $\bar{n}$ denote the mean photon number of the state. Probability of projection onto vacuum with an inefficient detector with efficiency $\eta$  will thus likely lie somewhere in between $\exp(-\eta\bar{n})$ and $1/(1+\eta\bar{n})$. In the asymptotic limit, when the state is almost perfectly Gaussified  and approaches thermal state, the interference of two identical thermal states at a balanced beam splitter will reproduce these thermal states at the output, c.f. Eq. (\ref{BSGtransform}). The asymptotic success probability of single iteration will therefore read  $1/(1+\eta \bar{n}_\infty)=(1-\eta|\alpha|^2)$, where  $\bar{n}_\infty$ is given by Eq. (\ref{nbarcohinfty}) and denotes the mean photon number of the asymptotic  thermal state.  Since $1-x^2\leq e^{-x^2}\leq 1/(1+x^2)$ we can conclude that the probability of each single iteration of the protocol $p_{\mathrm{succ},j}$ will lie in the interval from $1-\eta|\alpha|^2$ to $1/(1+\eta|\alpha|^2)$. These lower and upper bounds on sucess probabilities $p_{\mathrm{succ},j}$ are plotted in Fig.~\ref{figprobability} as solid and dashed black lines. }

\tcr{In contrast to iterative entanglement distillation protocol for two-qubit entangled states proposed by Bennett \emph{et al.} \cite{Bennett1996}, the asymptotic success probability of an elementary step of heralded Gaussification does not approach $1$ but is instead given by the vacuum probability of the (attenuated) asymptotic Gaussian state. Projections on other Gaussian states than vacuum can be considered to modify and improve the asymptotic success probability,  but this would  increase the experimental complexity. Moreover, in the LOCC settings one is restricted to local measurements, so projections onto entangled states are ruled out. }

\tcr{Previous realizations of heralded Gaussification \cite{Hage2008,Franzen2006,Hage2010} utilized homodyne measurements for heralding, instead of non-clicks of single-photon detectors, which further reduced the success probability of the protocol. The POVM $\hat{\Pi}_0$  defined in Eq. (9) can be implemented by performing eight-port homodyne detection which realizes generalized quantum measurement in the overcomplete basis of coherent states $|\alpha\rangle$. Each outcome $\alpha$ is  accepted with probability $P(\alpha)=\exp(-\frac{\eta}{1-\eta}|\alpha|^2)$ and rejected otherwise. The effective POVM element that corresponds to successful heralding reads
\begin{equation}
\hat{\Pi}_{\mathrm{EHD}}=\frac{1}{\pi}\int \exp\left(-\frac{\eta}{1-\eta}|\alpha|^2\right)  |\alpha\rangle\langle\alpha| d^2\alpha =(1-\eta)\sum_{n=0}^\infty (1-\eta)^n|n\rangle\langle n|=(1-\eta)\hat{\Pi}_0.
\end{equation}
With this implementation, the success probability of each heralding measurement is thus reduced by factor $1-\eta$ with respect to the scheme that utilizes single-photon detectors. If the homodyne detectors are not perfect and their detection efficiency is $\eta_{\mathrm{BHD}}<1$, then we must replace $\eta$ with $\eta/\eta_{\mathrm{BHD}}$ to achieve overall detection efficiency $\eta$ in the emulation of  $\hat{\Pi}_0$. Consequently, the success probability is reduced by factor $1-\eta/\eta_{\mathrm{BHD}}$. If $M$ heralding measurements  are performed simultaneously, the total reduction factor reads $(1-\eta/\eta_{\mathrm{BHD}})^M$. In our experiment, we achieve $\eta=0.4$ and $\eta_{\mathrm{BHD}}=0.65$, which yields success reduction factor $0.385$ for a single measurement and $0.057$ for three simultaneous conditioning measurements which are required for two iterations of heralded Gaussification. Conditioning on non-clicks of single-photon detectors thus provides significant advantage in terms of success probability of the protocol.}

\section{Summary and outlook}

In summary, we have experimentally investigated the convergence properties of heralded iterative Gaussification protocol.We have utilized phase-randomized coherent states with tunable mean photon number as convenient input states. We have implemented two iterations of the protocol and we have measured the input and the partially Gaussified states with a home-built homodyne detector. We have comprehensively characterized  the Gaussification procedure by analyzing the moments of quadrature distributions and photon number distributions and we have also reconstructed the full photon number distributions of the input and Gaussified states. We have compared the behavior of the heralded probabilistic Gaussification with a deterministic Gaussification where the conditioning measurement is absent. We have observed that the convergence of the heralded Gaussification rapidly slows down as we approach the convergence  threshold. Our results provide new insights into the behavior of the heralded Gaussification protocol and confirm the theoretical predictions concerning its convergence properties. 

\tcr{The main application of heralded Gaussification is the distillation of continuous-variable entanglement. By contrast, deterministic Gaussification cannot be used for entanglement distillation, because entanglement cannot be increased by local deterministic operations and classical communication. As illustrated and confirmed by our experiment, the speed of convergence of heralded Gaussification slows down significantly as we apporach the convergence threshold.  If highly entangled Gaussian states are targeted then many iterations of heralded Gaussification  may be required to fully distill  the entanglement. Also, even after several iterations of the protocol, the state may still detectably differ from a Gaussian state, which may play role in applications that rely on extremality of Gaussian states. }

\tcr{ Our experiment demonstrates the practical feasibility of heralded Gaussification based on heralding on non-clicks of single-photon detector which can  lead to significant improvement of the success probability of the protocol in comparison to approaches where the Gaussian measurements are performed with homodyne detectors.   For the phase-randomized input states considered in the present work, the Gaussification converges to thermal state. Therefore, our study is also connected to investigations of thermalization and it illustrates how conditional measurements (which can be also interpreted as measurements on the environment) can affect the thermalization process. }

In the future, the experiment can be extended in several ways. Superconducting single-photon detectors can be employed to increase the heralding detection efficiency. Implementation of third  iteration of the protocol can be attempted, which would require processing of eight copies of the input state. Other input non-Gaussian states can be considered, such as the phase-randomized squeezed states or mixtures of vacuum and single-photon states. In the latter case, conditionally generated single-photon states can be utilized, possibly subject to additional attenuation to control the ratio between the vacuum and single-photon components. Utilization of probabilistic sources of input states would greatly increase the complexity of the measurement and the required measurement time. These difficulties can be partially reduced by utilizing a time-multiplexing scheme where the Gaussified state circulates  in an optical  resonator and its interference with an additional input mode is switched on only when the latter is prepared in the desired state. 
Recently, it has been  demonstrated experimentally that such a quantum memory cavity \cite{Bouillard2019} can significantly increase the probability of breeding of larger Schr\"{o}dinger cat-like states \cite{Simon2024}, and this approach can be similarly beneficial also for heralded Gaussification.

\begin{backmatter}
\bmsection{Funding}
Czech Science Foundation (21-23120S), Univerzita Palack\'{e}ho v Olomouci (IGA-PrF-2024-008).

\bmsection{Disclosures}
The authors declare no conflicts of interest.

\bmsection{Data Availability Statement} 
Data underlying the results presented in this paper are not publicly available at this time but may be obtained from the authors upon reasonable request.

\end{backmatter}

\end{document}